\documentclass[11pt]{article}
\usepackage{latexsym}
\usepackage{amsfonts} 
\newtheorem{theorem}{Theorem}
\newtheorem{lemma}{Lemma}

\def\tr{{\rm tr}}

\begin{document}
\title{Existence of CMC and constant areal time foliations in $T^2$
  symmetric spacetimes with Vlasov matter} 
\author{H\aa kan Andr\'{e}asson\\Department of Mathematics\\
Chalmers University of Technology\\S-412 96 G\"{o}teborg, Sweden\\
hand@math.chalmers.se\\\phantom{hej}\\Alan D. Rendall\\
Max-Planck-Institut f\"{u}r Gravitationsphysik\\Am M\"{u}hlenberg 1,
D-14476 Golm, Germany\\rendall@aei-potsdam.mpg.de
\\\phantom{hej}\\Marsha Weaver\\Department of Physics, 
University of Alberta\\Edmonton, Alberta, Canada T6G 2J1\\mweaver@phys.ualberta.ca}

\date{} 
\maketitle 

\begin{abstract}

The global structure of solutions of the Einstein equations coupled to
the Vlasov equation is investigated in the presence of a two-dimensional 
symmetry group. It is shown that there exist global CMC and areal time
foliations. The proof is based on long-time existence theorems for the
partial differential equations resulting from the Einstein-Vlasov system
when conformal or areal coordinates are introduced.

\end{abstract} 

\section{Introduction}

In general relativity the gravitational field is expressed in terms of
the geometry of spacetime. The matter constituting self-gravitating physical
systems is described using certain matter fields which are also 
geometrical objects. The spacetime metric is required to satisfy the
Einstein equations and these are coupled to equations of motion for the
matter fields. The resulting Einstein-matter system of partial
differential equations is the central mathematical element of the theory.
Thus to understand the mathematical content of general relativity we 
require an overview of the solutions of these equations and their qualitative
behaviour.

When different kinds of physical situations are considered solutions of
the Einstein-matter equations with certain boundary conditions or spatial
asymptotics will be particularly relevant. One choice which avoids the issue 
of boundary conditions or asymptotic conditions in the strict sense is to
give initial data on a compact manifold. This possibility, which is that
studied in the present paper, is of interest for applications to cosmology.
In practice it leads to imposing periodic boundary conditions.

A further choice which has to be made in order to get a concrete mathematical
problem is to decide which kind of matter fields to consider. Here we choose
collisionless matter, where the equation of motion is the Vlasov equation.
The special interest of this kind of matter has been discussed in several
places (see \cite{r5}, \cite{r9}, \cite{a2}) and the relevant facts will not 
be repeated here. A fundamental insight is that collisionless matter acts
as a source of the gravitational field in a way which allows specific
features of the matter model to remain in the background while revealing 
basic properties of the dynamics of self-gravitating matter. A fact on the 
PDE level related to this is that in many cases long-time existence theorems
for solutions of the Einstein-Vlasov system are to be expected. For other
matter models coupled to the Einstein equations this is not true. Results
on formation of singularities in finite time in the case of one matter
model, dust, can be found in \cite{r2} and \cite{isenberg}. 

While the Einstein-matter system itself is naturally expressed in geometrical
terms the application of the theory of partial differential equations to 
study it requires the geometry to be parametrized in a suitable way by the
use of coordinates or other auxiliary constructs. An important step towards
investigating the dynamics of solutions is to find parametrizations which are
at once practical and applicable in sufficient generality. Of particular 
significance is finding a function which can act as a good time coordinate
in the situation of interest. This task is the focus of the following. It 
will be shown that under appropriate symmetry assumptions it can be solved 
in a very satisfactory way.

A central role in solving the geometric problems addressed in this paper
is played by two long-time existence results for certain systems of PDE in 
one space dimension. One of these is a global in time existence theorem
while the other is a continuation criterion which says that solutions
continue to exist as long as a certain quantity does not vanish. A number of 
techniques used in the proofs are adapted from known arguments for the 
Vlasov-Poisson and Vlasov-Maxwell systems while others are specific to the 
Einstein-Vlasov case.

Consider a solution of the Einstein-Vlasov system evolving from initial data
on a compact spacelike hypersurface. This paper is mainly concerned with the 
case where the initial hypersurface is a three-dimensional torus with periodic 
coordinates $(\theta, x, y)$. Moreover solutions are considered where both
the metric and the matter fields are invariant under translations in the 
$x$ and $y$ directions. Taking account of the periodic identifications
involved we see that these solutions admit a two-dimensional symmetry
group isomorphic to a two-torus $T^2=S^1\times S^1$. These spacetimes will 
be referred to as $T^2$ symmetric. No additional symmetry assumptions will
be imposed. Analogues of the results of this paper for various cases 
with higher symmetry, including certain subcases of $T^2$ symmetry, 
have been obtained previously. A survey of these earlier results can be found
in the introduction of \cite{arr}. The present paper can be seen as the 
culmination of a development concerning global geometrically defined time 
coordinates in solutions of the Einstein-Vlasov system with at least two 
symmetries. 

On a $T^2$ symmetric spacetime it is possible to define a function $R$,
the area function, as follows. If $p$ is a point of the spacetime then
$R(p)$ is equal to the area of the orbit of the action of $T^2$ which
contains $p$. Evidently the function $R$ is itself invariant under the
action of $T^2$. We say that a $T^2$ symmetric initial data set is {\it flat} 
if the spacetime gradient of $R$ in a Cauchy development of this data set 
vanishes identically on the initial hypersurface. Otherwise it will be called
non-flat. Whether an initial data set is flat in this sense can be determined 
intrinsically from the initial data without having to know anything about a 
Cauchy development. The reason for the terminology is that, as follows from 
\cite{r3}, if an initial data set is flat in this sense any Cauchy 
development of it is flat in the sense that its Riemann curvature tensor 
vanishes everywhere. Cf. the discussion in Section \ref{contract}.

The main theorems will now be stated. The first concerns the existence of
a global time coordinate of constant mean curvature. Consider a spacetime
which evolves from initial data on a compact hypersurface. In the following 
it will always be assumed implicitly that the initial datum for the particle
density has compact support. A real-valued function $t$ on the spacetime is 
called a constant mean curvature (CMC) time coordinate if each of its level 
hypersurfaces is compact and spacelike, it has constant mean curvature and  
the value of the mean curvature there is equal to $t$.

\begin{theorem}\label{theorem1}
Let $(M,g_{\alpha \beta},f)$ be the maximal globally hyperbolic 
development of non-flat $C^\infty$ initial data for the Einstein-Vlasov 
system with $T^2$ symmetry. Then $M$ can be covered by compact spacelike
hypersurfaces of constant mean curvature with each value in the range 
$(-\infty,0)$ occurring as the mean curvature of precisely one of these 
hypersurfaces.
\end{theorem}

The second theorem concerns the existence of an areal time coordinate. 
Consider a spacetime with $T^2$ symmetry. A real-valued function $t$ on the 
spacetime is called an areal time coordinate if each of its level 
hypersurfaces is compact and spacelike and the value of $t$ on the 
hypersurface is everywhere equal to that of the area function $R$.

\begin{theorem}\label{theorem2} 
Let $(M,g_{\alpha\beta},f)$ be the maximal globally hyperbolic development 
of non-flat $C^\infty$ initial data for the Einstein-Vlasov system with $T^2$ 
symmetry. Then $M$ can be covered by compact spacelike hypersurfaces of 
constant area function $R$ with each value in the range $(R_0,\infty)$ 
occurring as the value of the area function on precisely one of these 
hypersurfaces. Here $R_0$ is a non-negative real number.
\end{theorem}

Notice that these theorems include the vacuum case $f=0$. Theorem 2
was proved in the vacuum case in \cite{b1}. Under the additional
assumption of Gowdy symmetry, which consists in augmenting the
$T^2$ action by a suitable reflection symmetry, Theorem 2 was proved
in \cite{a}. In \cite{arr} an argument was sketched which indicates that 
Theorem 1 also holds under the assumption of Gowdy symmetry. The results 
of both theorems extend to the case of local $U(1)\times U(1)$ symmetry, 
as defined in \cite{r3}. Since there is no essential difference in the 
proofs this generalization will not be mentioned further. The fact that 
the theorems are stated for $C^\infty$ initial data is a matter of
convenience. Keeping track of derivatives would allow analogous
results to be proved for initial data of finite differentiability.

The paper is structured as follows. The next section contains some 
definitions which will be needed. The two sections after that contain the 
basic PDE analysis in the contracting and expanding directions respectively. 
The proofs of the main theorems are given in Section \ref{proof}. 
 
\section{The Einstein-Vlasov system with $T^2$ symmetry}\label{t2}
Consider the manifold $M=\mathbb{R}\times T^3$. Let $T^2$ 
act on $T^3$ in the obvious way, arising from the action of $\mathbb{R}^2$ on 
$\mathbb{R}^3$
with Cartesian coordinates $(\theta,x,y)$ by translations in $x$ and $y$. 
Correspondingly $T^2$ acts on $M$ with coordinates $(t,\theta,x,y)$.
A spacetime with underlying manifold $M$ defined by a metric 
$g_{\alpha\beta}$ and matter fields is said to be $T^2$ symmetric if the 
metric and matter fields are invariant under the action of $T^2$. The orbits 
of the group action will be referred to as surfaces of symmetry and a 
hypersurface will be called symmetric if it is a union of surfaces of 
symmetry. It is now clear how to define abstract Cauchy data for the 
Einstein-matter equations with $T^2$ symmetry. They should be defined on 
$T^3$ and invariant under the action of $T^2$. It will be assumed 
throughout that the metric and the matter fields are $C^\infty$.

Suppose now that a matter model is chosen for which the Cauchy problem for the 
Einstein equations is well-posed. The example of interest in
the following is that of collisionless matter satisfying the Vlasov 
equation. Corresponding to initial data for the Einstein-matter equations
with $T^2$ symmetry there is a maximal Cauchy development. We will construct 
a certain local coordinate system on a neighbourhood of the initial 
hypersurface in the maximal Cauchy development. On the initial hypersurface 
itself we choose periodic coordinates $(\theta,x,y)$ as above. The isometries 
are given by translations in $x$ and $y$. These coordinates can be extended
uniquely to a Gaussian coordinate system $(t,\theta,x,y)$ on a neighbourhood
of the initial hypersurface. On general grounds the action of $T^2$ on the
initial data extends uniquely to an action on the maximal Cauchy development 
$M$ by symmetries (see e.g. \cite{friedrich00}, Section 5.6). Gauss 
coordinates inherit the symmetries of their initial values and the spacetime. 
Hence the components of the metric in these coordinates depend only on 
$t$ and $\theta$. It is then a matter of simple algebra to see that in
these coordinates the metric can be written in the form
\begin{equation}
-dt^2+\mbox{e}^{2(\hat\eta-U)}d\theta^{2}
+\mbox{e}^{2U}[dx+Ady+(G+AH)d\theta]^{2}+
\mbox{e}^{-2U}R^{2}[dy+Hd\theta]^{2}.\label{datametric} 
\end{equation} 
for functions $(\hat\eta,U,A,G,H,R)$ of $t$ and $\theta$ which are periodic in 
$\theta$. It can also be read off that the function $R$ in this form of the
metric coincides with the area function mentioned in the introduction.

A region covered by coordinates of this type which is invariant
under the action of $T^2$ can be quotiented by the group action to
get a two dimensional quotient manifold $Q$ coordinatized by $t$ and 
$\theta$. The quotient inherits a Lorentzian metric by the rule that 
the inner product of two vectors on the quotient is equal to the
inner product of the unique vectors on spacetime which project onto
them orthogonally to the orbit. On $Q$ we can pass to double null 
coordinates $(u,v)$ on a neighbourhood of the quotient of the initial 
hypersurface, i.e. to coordinates whose level curves are null. Defining new
coordinates by $t=\frac{1}{2}(u-v)$ and $\theta=\frac{1}{2}(u+v)$ puts
the metric on $Q$ into conformally flat form. By pull-back these define
new coordinates on $M$ where the metric takes the form 
\begin{equation}\label{cmetric}
g=\mbox{e}^{2(\eta-U)}(-dt^{2}+d\theta^{2})+\mbox{e}^{2U}[dx+Ady
+(G+AH)d\theta]^{2}+
\mbox{e}^{-2U}R^{2}[dy+Hd\theta]^{2}.\label{confl} 
\end{equation} 
\phantom{hej}\\
for functions $(\eta,U,A,G,H,R)$ of $t$ and $\theta$.
A coordinate system of this type is called a conformal coordinate
system. It has now been seen that conformal coordinates always exist on 
some neighbourhood of the initial hypersurface in a spacetime evolving
from data with $T^2$ symmetry. In addition it is possible to choose the 
double null coordinates in such a way that the initial hypersurface coincides 
with $t=0$ and the metric coefficients are periodic in $\theta$.

To conclude this section we formulate the Einstein-Vlasov system which governs
the time evolution of a 
self-gravitating collisionless gas in the context of general relativity;
for the moment we do not assume any symmetry of the spacetime.
All the particles in the gas are assumed to have the same
rest mass, normalized to unity, and to move forward in time
so that their number density $f$ is a non-negative function
supported on the mass shell 
\[
PM := \left\{ \eta_{\mu \nu} v^\mu v^\nu = -1,\ v^0 >0 \right\},
\]
a submanifold of the tangent bundle $TM$ of the space-time manifold $M$
with metric $g_{\alpha \beta}$. Here $\eta_{\mu\nu}$ denotes the 
components of the Minkowski metric and $v^\mu$ denote the components of
a tangent vector in an orthonormal frame $e_\mu$. We use coordinates 
$(t,x^a)$ with zero shift and the frame components  $v^i$ to parametrize 
the mass shell; Greek indices always run from 0 to 3, and Latin ones from 
1 to 3. On the mass shell $PM$ the variable $v^0$ becomes a function of the 
$v^i$:
\[
v^0 = \sqrt{1+\delta_{ij}v^i v^j} .
\] 
The Einstein-Vlasov system now reads
\[
v^\mu  e_\mu(f)-\gamma^l_{\mu\nu}v^\mu v^\nu\frac{\partial f}{\partial v^l}=0
\]
\[
G^{\mu \nu} = 8 \pi T^{\mu \nu},
\]
\[
T^{\mu \nu}
= \int v^\mu v^\nu f \,\frac{dv^1 dv^2 dv^3}{-v_0}
\] 
where $\gamma^\lambda_{\mu \nu}$ are the Ricci rotation coefficients, 
$G^{\mu\nu}$ the Einstein tensor, and $T^{\mu\nu}$ 
the energy-momentum tensor. 
In the case of $T^2$ symmetry the number density 
$f$ is assumed to be invariant under the action induced on $PM$ by the action 
of $T^2$ on $M$. The orthonormal frame used to parametrize $PM$ in the 
following is also assumed to be invariant. As a result $f$ is independent of 
$x$ and $y$ and is a function of the variables $(t,\theta,v^1,v^2,v^3)$. An 
explicit choice of invariant orthonormal frame for the metric (\ref{cmetric}) 
is given by  
\[
e^{U-\eta}\frac{\partial}{\partial t}, e^{U-\eta}\left(\frac{\partial}{\partial
\theta}-G\frac{\partial}{\partial x}-H\frac{\partial}{\partial y}\right),
e^{-U}\frac{\partial}{\partial x},e^UR^{-1}
\left(\frac{\partial}{\partial y}-A\frac{\partial}{\partial x}\right)
\]

\section{Analysis in the contracting direction}\label{contract}

As we will see in more detail later, a non-flat $T^2$ symmetric solution of 
the Einstein-Vlasov system represents a cosmological model which, after a 
suitable choice of time orientation, evolves from an initial singularity
and has a phase of unlimited expansion at late times. In this section we 
consider the Einstein-Vlasov system in the contracting direction (i.e.
evolving towards the initial singularity) and use conformal coordinates in 
which the metric takes the form (\ref{cmetric}). That coordinates of this 
type can always be found near the initial hypersurface was shown in the last 
section. The invariant orthonormal frame adapted to these coordinates
exhibited there will be used in this section. The explicit form of the 
Einstein-Vlasov system will now be given. We set $\Gamma=G_t+AH_t$. The
quantities $\Gamma$ and $H_t$ will be referred to as {\it twist quantities}.
\phantom{hej}\\
\vspace{.2cm}
The Einstein-matter constraint equations
\begin{eqnarray}
&\displaystyle
U_{t}^{2}+U_{\theta}^{2}+\frac{\mbox{e}^{4U}}{4R^2}(A_{t}^2+A_{\theta}^2)
+\frac{R_{\theta\theta}}{R}-\frac{\eta_{t}R_{t}}{R}
-\frac{\eta_{\theta}R_{\theta}}{R}=&\nonumber\\
&\displaystyle =-\frac{\mbox{e}^{-2\eta+4U}}{4}\Gamma^2
-\frac{R^2\mbox{e}^{-2\eta}}{4}H_t^2
-\mbox{e}^{2(\eta-U)}\rho&\label{constr1con}\\
&\displaystyle
 2U_{t}U_{\theta}+\frac{\mbox{e}^{4U}}{2R^2}A_{t}A_{\theta}
+\frac{R_{t\theta}}{R}-\frac{\eta_{t}R_{\theta}}{R}
-\frac{\eta_{\theta}R_{t}}{R}=
\mbox{e}^{2(\eta-U)}J_1&\label{constr2con} 
\end{eqnarray} 
\phantom{hej}\\
\vspace{.2cm}
The Einstein-matter evolution equations
\begin{eqnarray} 
\displaystyle
U_{tt}-U_{\theta\theta}&=&\frac{U_{\theta}R_{\theta}}{R}
-\frac{U_{t}R_{t}}{R}+\frac{\mbox{e}^{4U}}{2R^2}(A_{t}^2-A_{\theta}^2)
+\frac{\mbox{e}^{-2\eta+4U}}{2}\Gamma^2
\nonumber\\ 
\displaystyle & &+\frac{1}{2}\mbox{e}^{2(\eta-U)}
(\rho-P_{1}+P_{2}-P_{3})\label{evol1con}\\ 
\displaystyle A_{tt}-A_{\theta\theta}&=&\frac{R_tA_t}{R}
-\frac{R_{\theta}A_{\theta}}{R}
+4(A_{\theta}U_{\theta}-A_tU_t)+R^2\mbox{e}^{-2\eta}\Gamma H_t\nonumber\\
\displaystyle & &+2R\mbox{e}^{2(\eta-2U)}S_{23}\label{evol4con}
\\ 
\displaystyle
R_{tt}-R_{\theta\theta}&=&R\mbox{e}^{2(\eta-U)}(\rho-P_{1})
+\frac{R\mbox{e}^{-2\eta+4U}}{2}\Gamma^2+\frac{R^3\mbox{e}^{-2\eta}}{2}H_t^2,
\label{evol2con}\\ 
\displaystyle\eta_{tt}-\eta_{\theta\theta}&=& 
U_{\theta}^{2}-U_{t}^{2}+\frac{\mbox{e}^{4U}}{4R^2}(A_{t}^2-A_{\theta}^2)
-\frac{\mbox{e}^{-2\eta+4U}}{4}\Gamma^2\nonumber\\ 
& &-\frac{3R^2\mbox{e}^{-2\eta}}{4}H_t^2-\mbox{e}^{2(\eta-U)}P_3
\label{evol3con} 
\end{eqnarray}
\phantom{hej}\\
\vspace{.2cm}
Auxiliary equations 
\begin{eqnarray}
\displaystyle &
&\partial_{\theta}[R\mbox{e}^{-2\eta+4U}\Gamma]
=-2R\mbox{e}^{\eta}J_2,\label{confaux1}\\
\displaystyle & & \partial_{t}[R\mbox{e}^{-2\eta+4U}\Gamma]
=2R\mbox{e}^{\eta}S_{12},\label{confaux2}\\
\displaystyle &
&\partial_{\theta}(R^3\mbox{e}^{-2\eta}H_t)
+R\mbox{e}^{-2\eta+4U}A_{\theta}\Gamma
=-2R^2\mbox{e}^{\eta-2U}J_3,\label{confaux3}\\ 
\displaystyle & &
\partial_{t}(R^3\mbox{e}^{-2\eta}H_t)+R\mbox{e}^{-2\eta+4U}A_{t}\Gamma
=2R^2\mbox{e}^{\eta-2U}S_{13}.\label{confaux4}
\end{eqnarray} 
\phantom{hej}\\
\vspace{.2cm}
The Vlasov equation
\begin{eqnarray}
&\displaystyle\frac{\partial f}{\partial
  t}+\frac{v^{1}}{v^{0}}\frac{\partial
  f}{\partial\theta}-\left[
(\eta_{\theta}-U_{\theta})v^{0}
+(\eta_{t}-U_{t})v^{1}-U_{\theta}\frac{(v^{2})^{2}}{v^{0}}\right.& 
\nonumber\\ 
&\displaystyle\left. +(U_{\theta}-\frac{R_{\theta}}{R})
\frac{(v^{3})^{2}}{v^{0}}-\frac{A_{\theta}}{R}\mbox{e}^{2U}
\frac{v^2v^3}{v^0}+\mbox{e}^{-\eta}(\mbox{e}^{2U}\Gamma
  v^2+RH_tv^3)
\right]
\frac{\partial f}{\partial v^{1}}\nonumber\\ 
&\displaystyle 
-\left[U_{t}v^{2}
+U_{\theta}\frac{v^{1}v^{2}}{v^{0}}
\right]\frac{\partial f}{\partial v^{2}}&
\nonumber\\ 
&\displaystyle -\left[(\frac{R_{t}}{R}-U_{t})v^{3}-(U_{\theta}
-\frac{R_{\theta}}{R})\frac{v^{1}v^{3}}{v^{0}}
+\frac{\mbox{e}^{2U}v^2}{R}(A_t+A_{\theta}\frac{v^1}{v^0})
\right]\frac{\partial f}{\partial v^{3}}=0.&\label{vvcon} 
\end{eqnarray}
The matter quantities
\begin{eqnarray} 
\rho(t,\theta)&=&\int_{\mathbb{R}^{3}}v^{0}f(t,\theta,v)\;dv\label{matt1}\\ 
P_{k}(t,\theta)&=&
\int_{\mathbb{R}^{3}}\frac{(v^{k})^{2}}{v^{0}}
f(t,\theta,v)\;dv,\;\;\;k=1,2,3\label{matt2}\\
J_k(t,\theta)&=&\int_{\mathbb{R}^{3}}v^{k}f(t,\theta,v)\;dv\label{matt3}\\ 
S_{jk}(t,\theta)&=&\int_{\mathbb{R}^{3}}\frac{v^jv^k}{v^0}f(t,\theta,v)\;dv
\label{matt4} 
\end{eqnarray}

Let a smooth $T^2$ symmetric solution of the Einstein-Vlasov system written
in conformal coordinates be given on some time interval $(t_-,t_0]$. We want 
to show that if this interval is bounded and if $R$ is bounded away from zero 
there then $f,\ R,\ \eta,\ U,\ A,\ G,\ H$ and all their derivatives are
bounded as well, with bounds depending on the data at $t=t_0$
and the lower bound on $R$, 
and that the supremum of the support of momenta at time $t$, 
\begin{equation}
Q(t):=\mbox{sup}\{|v|:\exists (s,\theta)\in [t,t_0]\times 
S^1\mbox{such that} f(s,\theta,v)\not= 0\}, 
\end{equation} 
is uniformly bounded. 
Note that these conditions imply that the matter quantities and 
their derivatives are uniformly bounded. In proving these statements it is 
assumed that the initial data at $t=t_0$ are non-flat, as in the assumptions 
of the main theorems.

The description of the proofs in this section is modelled on that of  
\cite{a} and highlights the places where there are differences due to
non-vanishing twist. 

\textit{Step 1. }(Monotonicity of $R$ and bounds on its first derivatives.)\\ 
This is a key step and follows the arguments in 
\cite{b1}. We have to check that the matter terms 
have the right signs so that these arguments still hold. The bounds 
on $R$ and its first derivatives will play a crucial role when we 
control the matter terms below. 

First we note that $\nabla R$ is timelike. This is a consequence of 
Proposition 3.1 of \cite{r3}. A $T^2$ symmetric solution of the 
Einstein-Vlasov system satisfies the hypotheses of that proposition.
Since the initial data set is non-flat the Hawking mass is non-zero
somewhere. Then the proposition implies that it is non-zero everywhere.
As a consequence $\nabla R$ is everywhere timelike.

Next we show that $\partial_{t}R$ and $|\partial_{\theta}R|$ are bounded 
into the past. Let us introduce the null vector fields 
\begin{equation}
\partial_{\xi}=\frac{1}{\sqrt{2}}(\partial_{t}+\partial_{\theta}),\;\;\;
\partial_{\lambda}=\frac{1}{\sqrt{2}}(\partial_{t}-\partial_{\theta}),
\label{null}
\end{equation}
and let us set $F_{\xi}=\partial_{\xi}F,\;F_{\lambda}=\partial_{\lambda}F$ 
for any function $F$. The evolution equation (\ref{evol2con}) can be written 
\begin{equation}\label{rxi}
\partial_{\lambda}R_{\xi}=\frac{R}{2}\mbox{e}^{2(\eta-U)}(\rho-P_1)
+\frac{R\mbox{e}^{-2\eta}}{4}(\mbox{e}^{4U}\Gamma^2+R^2H_t^2),\label{xilam}
\end{equation}
or equivalently, 
\begin{equation}\label{rlambda}
\partial_{\xi}R_{\lambda}=\frac{R}{2}\mbox{e}^{2(\eta-U)}(\rho-P_1)
+\frac{R\mbox{e}^{-2\eta}}{4}(\mbox{e}^{4U}\Gamma^2+R^2H_t^2).\label{lamxi}
\end{equation} 
The right hand side is positive since $\rho\geq P_1$ and we can conclude,
arguing as in Step 1 in Section 4 of \cite{a}, that 
both $R_t$ and $|R_{\theta}|$ are bounded into the past. 
Hence $R$ is uniformly $C^1$ bounded to the past of the initial surface. 

\textit{Step 2. }(Bounds on $U, A$ and $\eta$ and their first derivatives.)\\ 
The bounds on $U_{t},A_{t},U_{\theta}$ and $A_{\theta}$ to the past 
of the initial surface 
are obtained by a light-cone estimate, which in this case, with one spatial 
dimension, is an application of the Gronwall method on two independent 
null paths. Then, by combining these results, 
one obtains the desired estimate. 

The functions involved in the light-cone argument are quadratic functions 
in the first order derivatives of $U$ and $A,$ defined by 
\begin{eqnarray} 
X&=&\frac{1}{2}R(U_{t}^2+U_{\theta}^2)+\frac{\mbox{e}^{4U}}{8R}
(A_{t}^2+A_{\theta}^2),\label{X}\\ 
Y&=&RU_{t}U_{\theta}+\frac{\mbox{e}^{4U}}{4R}A_tA_{\theta}.\label{Y} 
\end{eqnarray} 
We will see below that if we let the vector fields along the null paths 
act on $X+Y$ and $X-Y$ we obtain 
equations appropriate for applying a 
Gronwall argument. 

Let us now derive bounds on $U$ and $A$ and their first order
derivatives. 
By using the evolution equations (\ref{evol1con}) and (\ref{evol4con}) 
we find 
\begin{eqnarray}
\displaystyle\partial_{\lambda}(X+Y)&=&\frac{-1}{2\sqrt{2}}R_{\xi}
\left(U_{t}^2-U_{\theta}^2+\frac{\mbox{e}^{4U}}{4R^2}
(-A_{t}^2+A_{\theta}^2)\right)
\nonumber\\
\displaystyle & &+\frac{R}{2}U_{\xi}\left(\mbox{e}^{-2\eta+4U}\Gamma^2+
\mbox{e}^{2(\eta-U)}(\rho-P_1+P_2-P_3)\right)
\nonumber\\
\displaystyle &
&+\frac{\mbox{e}^{2U}}{2R}A_{\xi}(R^2\mbox{e}^{2(U-\eta)}\Gamma H_t+
2R\mbox{e}^{2(\eta-U)}S_{23}),\nonumber 
\end{eqnarray}
and 
\begin{eqnarray} 
\displaystyle\partial_{\xi}(X-Y)&=&\frac{-1}{2\sqrt{2}}R_{\lambda}
\left(U_{t}^2-U_{\theta}^2+\frac{\mbox{e}^{4U}}{4R^2}
(-A_{t}^2+A_{\theta}^2)\right)
\nonumber\\
\displaystyle & &+\frac{R}{2}U_{\lambda} 
\left(\mbox{e}^{-2\eta+4U}\Gamma^2
+\mbox{e}^{2(\eta-U)}(\rho-P_1+P_2-P_3)\right) 
\nonumber\\ 
\displaystyle & &
+\frac{\mbox{e}^{2U}}{2R}A_{\lambda}(R^2\mbox{e}^{2(U-\eta)}\Gamma H_t+
2R\mbox{e}^{2(\eta-U)}S_{23}).\nonumber 
\end{eqnarray} 
It turns out that $X$ and $Y$ can be bounded by integrating these equations 
along null paths starting at a general point $(t_1,\theta)$ in the past of
the initial hypersurface and ending at the initial $t_0$ surface $t=t_0$. 
The boundedness of the integrals which arise follows from the equations
(\ref{rxi}) and (\ref{rlambda}) and the boundedness of $R_\xi$ and 
$R_\lambda$ in a similar way to the corresponding step in Section 4 of
\cite{a}. Up to multiplication by functions which are already bounded
and functions which can be bounded linearly in terms of $X$ 
the terms involving the twist whose integrals must be estimated are
$e^{-2\eta+4U}\Gamma^2$ and $e^{-2\eta+4U}\Gamma H_t$. The first of
these, multiplied by $R/4$, occurs as one of the positive summands on
the right hand side of (\ref{rxi}) and (\ref{rlambda}). Since by assumption
$R$ is bounded below the first term can be controlled. For the second
we can use the elementary inequality
\begin{equation}
e^{2(U-\eta)}\Gamma H_t\le \frac{1}{2}e^{-2\eta}(e^{4U}\Gamma^2+H_t^2)
\end{equation}    
and the occurrence of $\frac{R^3e^{-2\eta}}{4}H_t^2$ as a summand in 
(\ref{rxi}) and (\ref{rlambda}). Bounds on $X$ and $Y$ follow by applying
Gronwall's inequality as in \cite{a}. Thus, as long as $R$ stays uniformly 
bounded away from zero we conclude that $U$ and its first 
order derivatives, and thus also $A$ and its first order derivatives,
are bounded. Bounds on $|\eta|, |\eta_{t}|$ 
and $|\eta_{\theta}|$ are obtained in a similar way since the evolution 
equation (\ref{evol3con}) can be written 
\begin{equation} 
\displaystyle 2\partial_{\lambda}\eta_{\xi}=U_{\theta}^2-U_{t}^2
+\frac{\mbox{e}^{4U}}{4R^2}(A_{t}^2-A_{\theta}^2)
-\frac{\mbox{e}^{-2\eta+4U}}{4}\Gamma^2-\frac{3R^2\mbox{e}^{-2\eta}}{4}H_t^2
-\mbox{e}^{2(\eta-U)}P_{3},\nonumber 
\end{equation} 
or equivalently, 
\begin{equation} 
\displaystyle 2\partial_{\xi}\eta_{\lambda}=U_{\theta}^2-U_{t}^2
+\frac{\mbox{e}^{4U}}{4R^2}(A_{t}^2-A_{\theta}^2)
-\frac{\mbox{e}^{-2\eta+4U}}{4}\Gamma^2-\frac{3R^2\mbox{e}^{-2\eta}}{4}H_t^2
-\mbox{e}^{2(\eta-U)}
P_{3}.\nonumber 
\end{equation} 
The integrals of the right hand sides of these equations along null paths are
bounded since $P_3\le \rho-P_1$. The terms involving $\Gamma$ and $H_t$ 
can be handled as above. Thus
we find that 
$\eta$ is uniformly $C^1$ bounded to the past of the initial surface 
as long as $R$ stays bounded away from zero. 

\textit{Step 3. }(Bound on the support of the momentum.)\\ 
Note that a solution $f$ to the Vlasov equation is given by 
\begin{equation} 
f(t,\theta,v)=f_{0}(\Theta(0,t,\theta,v),V(0,t,\theta,v)),\label{solut} 
\end{equation} 
where $\Theta$ and $V$ are solutions to the characteristic system 
\begin{eqnarray} 
\displaystyle\frac{d\Theta}{ds}&=&\frac{V^1}{V^0}, 
\nonumber\\ 
\displaystyle\frac{dV^1}{ds}&=&-
(\eta_{\theta}-U_{\theta})V^{0}-(\eta_{t}-U_{t})V^{1}+U_{\theta}
\frac{(V^{2})^{2}}{V^{0}}
\nonumber\\
\displaystyle & &-(U_{\theta}-\frac{R_{\theta}}{R})\frac{(V^3)^2}{V^0}
+\frac{A_{\theta}}{R}
\mbox{e}^{2U}\frac{V^2V^3}{V^0} 
\nonumber\\ 
\displaystyle & &-\mbox{e}^{-\eta}(\mbox{e}^{2U}\Gamma V^2+RH_t V^3),
\nonumber\\ 
\displaystyle\frac{dV^2}{ds}&=&
-U_{t}V^{2}-U_{\theta}
\frac{V^{1}V^{2}}{V^{0}}, 
\nonumber\\ 
\displaystyle\frac{dV^3}{ds}&=&
-(\frac{R_{t}}{R}-U_{t})V^{3}
+(U_{\theta}-\frac{R_{\theta}}{R})\frac{V^{1}V^{3}}{V^{0}}
\nonumber\\ 
\displaystyle & &-\frac{\mbox{e}^{2U}
}{R}(A_t+A_{\theta}\frac{V^1}{V^0})V^2, 
\nonumber 
\end{eqnarray} 
and $\Theta(s,t,x,v),\;V(s,t,x,v)$ is the solution that goes 
through the point $(\theta,v)$ at time $t$. 
Let us recall the definition of 
$$Q(t):=\mbox{sup}\{|v|:\exists (s,\theta)\in [t,t_0]\times
S^1\mbox{such that} f(s,\theta,v)\not= 0\}.$$ We also define
$Q^j,\;j=1,2,3$ in the obvious way where $|v|$ is replaced by $|v^j|.$ 
If $Q(t)$ can be controlled we
obtain immediately from (\ref{matt1})-(\ref{matt3}) bounds
on $\rho, J_k,P_k,$ and $S_{jk},$ \, $j,k=1,2,3$, $j\ne k$,
since $\|f\|_{\infty}\leq
\|f_0\|_{\infty}$ from (\ref{solut}). 
First we note that the quantities $$\mbox{e}^{U}V^2,\; 
A\mbox{e}^{U}V^2+R\mbox{e}^{-U}V^3$$ are conserved which a simple 
computation shows by using the equations for $dV^2/ds$ and $dV^3/ds$
above. More generally this is a consequence of the fact that if
$\gamma$ is a geodesic and $k$ a Killing field then $g(\gamma',k)$ is
conserved along the geodesic. Here $\gamma'$ is the tangent vector
to $\gamma$ and since the particles follow the geodesics of spacetime
the tangent vector can be expressed in terms of
$v^\mu.$ We chose $k=\partial_x$ and $k=\partial_y$ as
Killing fields to derive the conserved quantities given above. 
Since $U$ and $A$ are uniformly bounded as long as $R$ stays bounded
away from zero we conclude that $V^2$ and $V^3$ and thus $Q^2$ and
$Q^3$ are bounded as well. 
In order to bound $Q(t)$ we need to control the remaining component $Q^1$. 
From the auxiliary equations (\ref{confaux2}) and (\ref{confaux4}) we 
conclude that
$\Gamma$ and $H_t$ can be bounded by the quantities involving 
$\eta, U, A_t,$ which are known to be bounded, and by $|S_{12}|$ and
$|S_{13}|.$ Since $Q^2$ and $Q^3$ are controlled we immediately get
from (\ref{matt4}) that $|S_{12}|$ and $|S_{13}|$ are bounded by a
constant times $Q^1.$ We conclude that as long as $R$ stays bounded
away from zero the terms involving $\Gamma$ and $H_t$ in the
characteristic equation for $dV^1/ds$ can be estimated by
$C(t)Q^1(t),$ where $C(t)$ is uniformly bounded on closed time
intervals. Now, since the field components $U,A$ and $\eta$ and their first 
derivatives are known to be
bounded on $(t_-,t_0]$ (as long as $R$ stays bounded away from
zero) we obtain from the characteristic equation for $V^1$ 
\begin{displaymath}
|V^1(t)|\leq |V^1(t_0)|
+C(t)\int_{t}^{t_0}Q^1(s)ds
\leq |Q^1(t_0)|+C(t)\int_{t}^{t_0}Q^1(s)ds,\; t<t_0. 
\end{displaymath} 
Note that $Q^1(t_0)$ is bounded by a positive constant since $f_0$ has
compact support. 
This
inequality leads to a Gronwall inequality for $Q^1(t)$ and we conclude
that $Q^1(t)$ is uniformly bounded on $[t,t_0].$ 

Thus all the field components, their first derivatives and the matter
terms are known to be 
bounded on $(t_-,t_0]$, as long as $R$ stays bounded away from zero. 

\textit{Step 4. }(Bounds on the second order derivatives of the field 
components and on the first order derivatives of $f$.) 
From the Einstein-matter constraint equations in conformal coordinates 
we can express $R_{t\theta}$ and $R_{\theta\theta}$ in terms of 
uniformly bounded 
quantities, as long as $R$ stays bounded away from zero. Therefore these 
functions are uniformly bounded and 
equation (\ref{evol2con}) then implies that $R_{tt}$ is uniformly 
bounded as well. 

In the vacuum case one can take the derivative of the evolution equations and 
repeat the argument in Step 2 to obtain bounds on second order derivatives 
of $U$ and $A$. Here we need another argument. 
First we write the evolution equations for $U$ 
and $A$ in the forms 
\begin{eqnarray}
\displaystyle U_{tt}-U_{\theta\theta}&=&
\frac{(R_{\theta}-R_{t})}{2R}(U_{\theta}+U_{t})
-\frac{(R_{\theta}+R_{t})}{2R}(U_{t}-U_{\theta}) 
\nonumber\\ 
\displaystyle & &+\frac{\mbox{e}^{4U}}{2R^2}(A_t-A_{\theta})(A_t+A_{\theta}) 
+\frac{\mbox{e}^{-2\eta+4U}}{2}\Gamma^2+\frac{1}{2}\mbox{e}^{2(\eta-U)}\kappa, 
\nonumber 
\end{eqnarray} 
and 
\begin{eqnarray} 
\displaystyle A_{tt}-A_{\theta\theta}&=&\frac{(R_{t}-R_{\theta})}{2R}
(A_{\theta}+A_{t})
+\frac{(R_{\theta}+R_{t})}{2R}(A_{t}-A_{\theta})
\nonumber\\ 
\displaystyle & &-2(A_{t}-A_{\theta})(U_{\theta}+U_{t})
-2(A_{\theta}+A_{t})(U_{t}-U_{\theta})
\nonumber\\ 
\displaystyle & &+R^2\mbox{e}^{-2\eta}\Gamma H_t 
+2R\mbox{e}^{2(\eta-2U)}S_{23},\nonumber 
\end{eqnarray} 
where $\kappa$ denotes $\rho-P_1+P_2-P_3$. 
Taking the $\theta$-derivative of these equations gives 
\begin{eqnarray} 
\partial_{\lambda}\partial_{\xi}U_{\theta}&=&L+\frac{R_{\lambda}}{2R}
\partial_{\xi}
U_{\theta}+ 
\frac{R_{\xi}}{2R}\partial_{\lambda}U_{\theta}
+\frac{\mbox{e}^{4U}}{2R^2}
(A_{\lambda}\partial_{\xi}A_{\theta}+A_{\xi}\partial_{\lambda}A_{\theta}) 
\nonumber\\ 
\displaystyle & &
+\frac{1}{2}(\mbox{e}^{-2\eta+4U}\Gamma_{\theta}
+\mbox{e}^{2(\eta-U)}\kappa_{\theta}),\label{dthe1} 
\end{eqnarray} 
and 
\begin{eqnarray} 
\partial_{\lambda}\partial_{\xi}A_{\theta}&=&L+\frac{R_{\lambda}}{2R}
\partial_{\xi}
A_{\theta}-\frac{R_{\xi}}{2R}\partial_{\lambda}A_{\theta}+ 
2U_{\xi}\partial_{\lambda}A_{\theta}+2A_{\lambda}\partial_{\xi}U_{\theta}
+2U_{\lambda}\partial_{\xi}A_{\theta}
\nonumber\\ 
\displaystyle & &
+2A_{\xi}\partial_{\lambda}U_{\theta}
+R^2\mbox{e}^{-2\eta}(\Gamma H_t)_{\theta} 
+2R\mbox{e}^{2(\eta-2U)}(S_{23})_{\theta}, 
\end{eqnarray} 
Here, $L$ contains only $\Gamma, H_t,\kappa$ and $S_{23}$, 
first order derivatives of $U, A$ and $\eta$, and first and second order 
derivatives of $R$, which all are known to be bounded. 
These equations can of course also be written in a form where 
the left hand sides read $\partial_{\xi}\partial_{\lambda}U_{\theta}$ 
and $\partial_{\xi}\partial_{\lambda}A_{\theta},$ respectively. 
By integrating these equations along null paths to the past of the initial 
surface, we get from a Gronwall argument a bound 
on $$\sup_{\theta\in S^1}(|\partial_{\xi}U_{\theta}|
+|\partial_{\lambda}U_{\theta}|+|\partial_{\xi}A_{\theta}|
+|\partial_{\lambda}A_{\theta}|),$$ as long 
as $R$ is bounded away from zero, under the hypothesis that the integral 
of the differentiated terms $\Gamma_{\theta},H_{t\theta},
\kappa_{\theta}$ and $(S_{23})_{\theta}$ 
can be controlled. That the first two terms are bounded follows
immediately from the auxiliary equations (\ref{confaux1}) and
(\ref{confaux3}) in view of the bound on $Q(t).$ 
In order to bound the matter terms we make use of a device introduced 
by Glassey and Strauss \cite{g1} for treating the Vlasov-Maxwell equation.
Since there is no essential difference to the case treated in \cite{a}
we do not repeat the proof here. This procedure allows the integrals of 
the differentiated matter terms to be controlled and the Gronwall argument 
referred to above goes through. So we obtain uniform bounds 
on $|\partial_{\xi}U_{\theta}|, |\partial_{\lambda}U_{\theta}|, 
|\partial_{\xi}A_{\theta}|$, and $|\partial_{\lambda}A_{\theta}|,$ and 
therefore also on $|U_{\theta\theta}|, |U_{t\theta}|, |A_{\theta\theta}|$ 
and $|A_{t\theta}|,$ as long as $R$ is 
bounded away from zero. The evolution 
equations (\ref{evol1con}) and (\ref{evol4con}) then give uniform bounds 
on $|U_{tt}|$ and $|A_{tt}|$. Bounds on second order derivatives of
$G$ and $H$ follow from the auxiliary equations. 
By differentiating equation (\ref{evol3con}), it is now straightforward to 
obtain bounds on the second order derivatives of $\eta$, using similar 
arguments to those already discussed here, in particular the integrals 
involving matter quantities can be treated as above. Bounds 
on the first order derivatives of the distribution function $f$ may now be 
obtained from the known bounds on the field components from the formula 
\begin{equation}
f(t,\theta,v)=f_{0}(\Theta(0,t,\theta,v),V(0,t,\theta,v)), 
\end{equation}\\ 
since $f_0$ is smooth and since $\partial\Theta$ and 
$\partial V$ (here $\partial$ denotes $\partial_{t}, 
\partial_{\theta}$ or $\partial_v$) can be controlled by a Gronwall argument 
in view of the characteristic system and the auxiliary equations
(\ref{confaux1})-(\ref{confaux4}). 

\textit{Step 5. }(Bounds on higher order derivatives and completion 
of the proof.) 
It is clear that the method described above can be continued for 
obtaining bounds on higher derivatives as well. Hence, we have uniform bounds 
on the functions $R,U,A,\eta$ and $f$ and all their derivatives 
on the interval $(t_-,t_0]$ if $R>\epsilon>0$. 
This implies that the solution 
extends to $t\rightarrow -\infty$ as long as $R$ stays bounded away from zero. 

Later we will require a slight generalization of these results in order to
show that the arguments of Section 5 of \cite{arr} generalize to cover the
case of $T^2$ symmetry. Once it has 
been established that $\nabla R$ is timelike, the estimates in the later 
steps hold for 
any subset $Z$ of the half-plane $t\le t_0$ provided $Z$ is a future set. 
By definition this means that any future directed causal curve in the 
region $t\le t_0$ starting at a point of $Z$
remains in $Z$. (For information on concepts such as this 
concerning causal structures see e.g. \cite{h1}.) Thus if $R$ is 
bounded away from zero on $Z$ and $t$ is bounded on $Z$ then all the 
unknowns and their derivatives can be controlled on $Z$.

Now consider a special choice of the subset $Z$, namely that which is 
defined by the inequalities $t_1<t\le t_0$ and 
$\theta_1+t_0-t<\theta<\theta_2-t_0+t$ for some numbers $\theta_1$, 
$\theta_2$ and 
$t_1$ satisfying the inequalities $\theta_1<\theta_2$ and $t_1>t_0-(1/2)
(\theta_2-\theta_1)$.
Suppose a solution of the Einstein-Vlasov system in conformal coordinates
defined on this region is such that $R$ is bounded away from zero. Then the
functions defining the solution extend smoothly to the boundary of $Z$ at
$t=t_1$. They define smooth Cauchy data for the Einstein-Vlasov system.
Applying the standard local existence theorem (without symmetry) allows 
the solution to be extended through that boundary. Repeating the 
construction of conformal coordinates in Section~\ref{t2} then
shows that we get an extension of the solution written in conformal 
coordinates through that boundary.

\section{Analysis in the expanding direction}\label{expand}

In this section we want to investigate the Einstein-Vlasov
system with $T^2$ symmetry in the expanding direction.
We write the system in areal coordinates, i.e., the coordinates are
chosen such that $R=t$. The circumstances under which coordinates of this
type exist are discussed in Section \ref{proof}.
We prove that for initial data on 
a hypersurface of constant time corresponding solutions exist for all 
future time with respect to the areal time coordinate. In order to extend
a solution defined on a finite time interval in these coordinates  
to one which exists globally in time it is sufficient to obtain uniform bounds 
on the field components and the distribution function and all 
their derivatives on a finite time interval $[t_0,t_+)$ on 
which the local solution exists. For in this case the functions defining
the solution are uniformly continuous and extend continuously to $t=t_+$.
Since all derivatives are bounded the extension is smooth and we obtain new
initial data on $t=t_+$. Applying the general local existence theorem for the
Einstein-Vlasov system \cite{c1}
allows the spacetime to be extended beyond $t=t_+$.
Conformal coordinates can be introduced on a neighbourhood of $t=t_+$ and
from those areal coordinates can be constructed. In this way an extension
to the future of the solution written in areal coordinates is obtained. 

Below the form of the metric and the Einstein-Vlasov system are given in 
areal coordinates. The functions $\alpha,\eta,U,A,G,H$ all depend on $t$ 
and $\theta$ and the function $f$ depends on $t,\theta$ and 
$v\in\mathbb{R}^3$ and as before we have set $\Gamma=G_t+AH_t$.
The orthonormal frame used to parametrize $PM$ is
\[
\alpha^{1/2}e^{U-\eta}\frac{\partial}{\partial t}, 
e^{U-\eta}\left(\frac{\partial}{\partial
\theta}-G\frac{\partial}{\partial x}-H\frac{\partial}{\partial y}\right),
e^{-U}\frac{\partial}{\partial x},e^Ut^{-1}
\left(\frac{\partial}{\partial y}-A\frac{\partial}{\partial x}\right)
\]
\vspace{.3cm}
Metric 
\begin{eqnarray} 
g&=&\mbox{e}^{2(\eta-U)}(-\alpha dt^{2}+
d\theta^{2})+\mbox{e}^{2U}[dx+Ady+(G+AH)d\theta]^{2}\nonumber\\ 
& &+\mbox{e}^{-2U}t^{2}[dy+Hd\theta]^{2}\label{areal} 
\end{eqnarray} 
\phantom{hej}\\
\vspace{.3cm}
The Einstein-matter constraint equations
\begin{eqnarray}
\displaystyle\frac{\eta_{t}}{t}&=&U_{t}^{2}+\alpha U_{\theta}^{2}+
\frac{\mbox{e}^{4U}}{4t^2}(A_{t}^2+\alpha
A_{\theta}^2)+\frac{\mbox{e}^{-2\eta}}{4}
(\mbox{e}^{4U}\Gamma^2+t^2H_t^2)\nonumber\\
& &+\mbox{e}^{2(\eta-U)}\alpha\rho\label{constr1}\\
\displaystyle\frac{\eta_{\theta}}{t}&=&2U_{t}U_{\theta}+
\frac{\mbox{e}^{4U}}{2t^2}
A_{t}A_{\theta}-\frac{\alpha_{\theta}}{2t\alpha}
-\mbox{e}^{2(\eta-U)}\sqrt{\alpha}J_1\label{constr2}\\
\displaystyle\alpha_{t}&=&2t\alpha^{2}\mbox{e}^{2(\eta-U)}(P_{1}-\rho)-
\alpha t\mbox{e}^{-2\eta}(\mbox{e}^{4U}\Gamma^2+t^2H_t^2)
\label{constr3} 
\end{eqnarray} 
\phantom{hej}\\ 
\vspace{.3cm}
The Einstein-matter evolution equations 
\begin{eqnarray} 
\displaystyle\eta_{tt}
-\alpha\eta_{\theta\theta}&=&\frac{\eta_{\theta}\alpha_{\theta}}{2}+
\frac{\eta_{t}\alpha_{t}}{2\alpha}
-\frac{\alpha_{\theta}^{2}}{4\alpha}
+\frac{\alpha_{\theta\theta}}{2}-U_{t}^{2}+\alpha
U_{\theta}^{2}\nonumber\\
\displaystyle & &+\frac{\mbox{e}^{4U}}{4t^2}(A_{t}^2-\alpha A_{\theta}^2)
-\frac{3}{4}\mbox{e}^{-2\eta}t^2H_t^2
-\frac{1}{4}\mbox{e}^{-2\eta}\mbox{e}^{4U}\Gamma^2\nonumber\\ 
& &-\alpha\mbox{e}^{2(\eta-U)}P_3, 
\label{evol1}\\ 
\displaystyle U_{tt}-\alpha U_{\theta\theta}&=&
-\frac{U_{t}}{t}+\frac{U_{\theta}\alpha_{\theta}}{2}
+\frac{U_{t}\alpha_{t}}{2\alpha}+
\frac{\mbox{e}^{4U}}{2t^2}(A_{t}^2-\alpha A_{\theta}^2)\nonumber\\
\displaystyle & &+\frac{\mbox{e}^{-2\eta}\mbox{e}^{4U}}{2}\Gamma^2 
+\frac{\mbox{e}^{2(\eta-U)}\alpha}{2}(\rho-P_{1}+P_{2}-P_{3})
\label{evol2}\\
\displaystyle A_{tt}-\alpha A_{\theta\theta}&=&\frac{A_t}{t}+
\frac{\alpha_{\theta}A_{\theta}}{2}+\frac{\alpha_{t}A_{t}}{2\alpha}-4A_tU_t
+4\alpha A_{\theta}U_{\theta}\nonumber\\
\displaystyle & &+t^2\mbox{e}^{-2\eta}\Gamma H_t 
+2t\alpha\mbox{e}^{2(\eta-2U)}S_{23}.\label{evol4} 
\end{eqnarray} 
\phantom{hej}\\
\vspace{.3cm}
Auxiliary equations 
\begin{eqnarray}
\displaystyle 
\partial_{\theta}[\mbox{e}^{-2\eta}\alpha^{-1/2}\mbox{e}^{4U}\Gamma ]
 &=&-2\mbox{e}^{\eta}J_{2},\label{aux1}\\
\displaystyle
\partial_{t}[\mbox{e}^{-2\eta}t\alpha^{-1/2}\mbox{e}^{4U}\Gamma ] &=&
2t\alpha^{1/2}\mbox{e}^{\eta}S_{12},\label{aux2}\\
\displaystyle 
\partial_{\theta}[\mbox{e}^{-2\eta}\alpha^{-1/2}(A\mbox{e}^{4U}\Gamma 
+t^{2}H_t)]
 &=&-2\mbox{e}^{\eta}AJ_2-2t\mbox{e}^{\eta-2U}J_{3},\label{aux3}\\ 
\displaystyle
\partial_{t}[\mbox{e}^{-2\eta}t\alpha^{-1/2}(A\mbox{e}^{4U}\Gamma +t^{2}H_t)]
&=&2t\alpha^{1/2}\mbox{e}^{\eta}(AS_{12}+t\mbox{e}^{-2U}S_{13}).\label{aux4}
\end{eqnarray}
\phantom{hej}\\
\vspace{.3cm}
The Vlasov equation
\begin{eqnarray}
&\displaystyle\frac{\partial f}{\partial
  t}+\frac{\sqrt{\alpha}v^{1}}{v^{0}}\frac{\partial
  f}{\partial\theta}-\left[
(\eta_{\theta}-U_{\theta}+\frac{\alpha_{\theta}}{2\alpha})\sqrt{\alpha}v^{0}
+(\eta_{t}-U_{t})v^{1}
\right.& 
\nonumber\\
&\displaystyle\left.
-\frac{\sqrt{\alpha}\mbox{e}^{2U}A_{\theta}}{t}
\frac{v^2v^3}{v^0}+\frac{\sqrt{\alpha}U_{\theta}}{v^{0}}((v^{3})^{2}
-(v^{2})^{2})+\mbox{e}^{-\eta}(\mbox{e}^{2U}\Gamma v^2
+tH_tv^3)\right]\frac{\partial
  f}{\partial v^{1}}\nonumber\\ 
&\displaystyle
-\left[U_{t}v^{2}+\sqrt{\alpha}U_{\theta}\frac{v^{1}v^{2}}{v^{0}}\right]
\frac{\partial f}{\partial v^{2}}& 
\nonumber\\
&\displaystyle-\left[(\frac{1}{t}-U_{t})v^{3}
-\sqrt{\alpha}U_{\theta}\frac{v^{1}v^{3}}{v^{0}}+\frac{\mbox{e}^{2U}v^2}{t}
(A_t+\sqrt{\alpha}A_{\theta}\frac{v^1}{v^0})\right]\frac{\partial
  f}{\partial v^{3}}=0.&\label{vv} 
\end{eqnarray} 
\phantom{hej}\\
The matter quantities are defined as in (\ref{matt1})-(\ref{matt4}).

\textit{Step 1. }(Bounds on $\alpha, U, A,G,H$ and $\tilde{\eta}$.)\\ 
In this step we first show an ``energy'' monotonicity lemma and then 
we show how this result leads to bounds on $\tilde{\eta}:=\eta+\ln{\alpha}/2$ 
and on $U$ and $A$. 
Let $E(t)$ be defined by 
\begin{eqnarray}
\displaystyle E(t)&=&\int_{S^1}\left[\alpha^{-\frac{1}{2}}U_{t}^2+\sqrt{\alpha}
U_{\theta}^2+\frac{\mbox{e}^{4U}}{4t^2}(\alpha^{-\frac{1}{2}}A_{t}^2
+\sqrt{\alpha}
A_{\theta}^2)\right.\nonumber\\ 
\displaystyle & 
&\left.+\frac{\mbox{e}^{-2\eta}\alpha^{-1/2}}{4}(\mbox{e}^{4U}\Gamma^2
+t^2H_t^2)+\sqrt{\alpha}\mbox{e}^{2(\eta-U)}\rho\right]
  d\theta.
\end{eqnarray}
\begin{lemma} 
$E(t)$ is a monotonically decreasing function in $t$, and satisfies
\begin{eqnarray} 
\displaystyle \frac{d}{dt}E(t)&=&-\frac{2}{t}\int_{S^1}
\left[\frac{U_{t}^2}{\sqrt{\alpha}} 
+\frac{\mbox{e}^{4U}}{4t^2}\sqrt{\alpha}A_{\theta}^2
+\frac{\mbox{e}^{-2\eta}}{4\sqrt{\alpha}}(\mbox{e}^{4U}\Gamma^2
+2t^2H_t^2)\right.\nonumber\\ 
\displaystyle & &\left. +\frac{\sqrt{\alpha}}{2} 
\mbox{e}^{2(\eta-U)}(\rho+P_3)\right] d\theta\leq 0.\label{energ} 
\end{eqnarray} 
\end{lemma} 
\textit{Proof. }This is a straightforward but lengthy
computation. Let us sketch the steps involved. After taking the time
derivative of the integrand we use the evolution equations for $U$ and
$A$ to substitute for the seond order derivatives, we use the
auxiliary equations to express second order derivatives of $G$ and $H$
in terms of matter quantities and we express $\rho_t$ by using the
Vlasov equation. Integrating by parts and using the constraint
equations for $\eta_t$ and $\alpha_t$ lead to (\ref{energ}). 
\begin{flushright} 
$\Box$ 
\end{flushright} 

Let us now define the quantity $\tilde{\eta}$ by 
\begin{equation} 
\tilde{\eta}=\eta+\frac{1}{2}\ln{\alpha}. 
\end{equation} 
The difference $\tilde\eta(t,\theta_1)-\tilde\eta(t,\theta_2)$ can be 
estimated by integrating the expression for $\tilde\eta_\theta$ 
resulting from (\ref{constr2}) and using the energy bound. Since 
(\ref{constr2})
does not contain the twist quantities this is exactly as in \cite{a}.
The next step is to use the constraint equations (\ref{constr1}) and 
(\ref{constr3}) to bound the integral $\int_{S^1}\tilde\eta d\theta$.
The net contribution to $\tilde\eta_t$ from the twist quantities has
the opposite sign from that of the other terms. So when obtaining an
upper bound for the integral it can be discarded and the argument proceeds
as in \cite{a}. On the other hand the twist terms must be taken into
account when obtaining a lower bound for the integral. The expression which
has to be estimated is equal to the twist contribution to the energy up
to a factor $\alpha$. This factor is bounded since $\alpha$ is monotone 
decreasing. Knowing that the difference of $\tilde\eta$ at two points 
$\theta_1$
and $\theta_2$ at any given time and its integral at any given time are
bounded it follows that $\tilde\eta(t,\theta)\le C(t)$ for some bounded 
function $C(t)$

\textbf{Remark. }In the analysis below $C(t)$ will always denote a 
uniformly bounded function on $[t_0,t_+)$. Sometimes we introduce other 
functions with the same property only for the purpose of trying to make 
some estimates become more transparent.\\ 

Next we show that the boundedness of $E(t)$, 
together with the constraint equation (\ref{constr3}), lead 
to a bound on $|U|$. 
For any $\theta_1,\theta_2\in S^1$, and $t\in [t_0,t_+)$
we get by H\"{o}lder's inequality 
\begin{eqnarray} 
&\displaystyle|U(t,\theta_2)-U(t,\theta_1)|=\left|\int_{\theta_1}^{\theta_2}
U_{\theta}(t,\theta)d\theta\right|&
\nonumber\\ 
&\displaystyle\leq\left(\int_{\theta_1}^{\theta_2}\alpha^{-1/2}d\theta\right)
^{1/2}\left(\int_{\theta_1}^{\theta_2}\sqrt{\alpha}U_{\theta}^2d\theta\right)
^{1/2}.&\label{Hlr} 
\end{eqnarray} 
The second factor on the right hand side is clearly 
bounded by $(E(t_0))^{1/2}$. For the first factor we use the constraint 
equation (\ref{constr3}). 
This equation can be written as 
\begin{equation} 
\partial_{t}(\alpha^{-1/2})=t\sqrt{\alpha}\mbox{e}^{2(\eta-U)}(\rho-P_1)
+t\alpha^{-1/2}
\frac{\mbox{e}^{-2\eta}}{2}(\mbox{e}^{4U}\Gamma^2+t^2H_t^2)
, 
\end{equation} 
so that for $t\in [t_0,t_+)$ 
\begin{eqnarray} 
\displaystyle\alpha^{-1/2}(t,\theta)&=&\int_{t_0}^{t}s\sqrt{\alpha}
\mbox{e}^{2(\eta-U)}(\rho-P_1)+s\alpha^{-1/2} 
\frac{\mbox{e}^{-2\eta}}{2}(\mbox{e}^{4U}\Gamma^2+s^2H_t^2)ds\nonumber\\
\displaystyle & & 
+\alpha^{-1/2}(t_0,\theta). 
\end{eqnarray} 
Since $\rho\geq P_1$, the integrand is positive and bounded by a multiple
of the integrand of $E(t)$. 
Letting $C$ denote the supremum 
of $\alpha^{-1/2}(t_0,\cdot)$ over $S^1$ we get 
\begin{displaymath} 
\int_{\theta_1}^{\theta_2}\alpha^{-1/2}d\theta\leq\int_{t_0}^{t}2sE(s) ds
+2\pi C 
\leq 2E(t_0)(t^2-t_{0}^2)/2+2\pi C. 
\end{displaymath} 
Hence, for any $\theta_1,\theta_2\in S^1$ we have 
\begin{equation} 
|U(t,\theta_2)-U(t,\theta_1)|\leq C(t). 
\end{equation} 
Next $\int_{S^1}U(t,\theta)d\theta$ can be estimated using the energy
just as in \cite{a} 
since the twist quantities do not play a role in that argument. Knowing
that the difference of $U$ at any two spatial points and the modulus of
its integral over the circle are bounded we can conclude that $U$ itself 
is bounded. These arguments also apply
to $A$, since the factor $\mbox{e}^{4U}$ is controlled by the uniform bound 
on $U$. Bounds on $\mbox{e}^{-\eta}\Gamma$ and $\mbox{e}^{-\eta}H_t$ 
also follow from these arguments. 
Indeed, let 
$P=\mbox{e}^{4U-2\eta}\alpha^{-1/2}\Gamma.$ We have from the auxiliary
equation (\ref{aux1}) an expression for $P_{\theta}$ in terms of $J_2$
and we get 
\begin{eqnarray} 
&\displaystyle|P(t,\theta_2)-P(t,\theta_1)|=\left|\int_{\theta_1}^{\theta_2}
P_{\theta}(t,\theta)d\theta\right|
\leq\int_{\theta_1}^{\theta_2}2\mbox{e}^{\eta}|J_2|
d\theta&
\nonumber\\ 
&\displaystyle 
\leq 2\|\mbox{e}^{-\tilde{\eta}+2U}\|_{\infty}
\int_{\theta_1}^{\theta_2}\sqrt{\alpha}\mbox{e}^{2(\eta-U)}|J_2|d\theta\leq
C(t)E(t).& 
\end{eqnarray} 
Note that $\tilde{\eta}$ and $U$ are known to be bounded. Similarily,
using the auxiliary equation (\ref{aux2}) we obtain a bound on
$$\left|\int_{S^1}P(t,\theta)d\theta\right|$$ in the same spirit as
for $U.$ This leads to a bound on $P$ itself which implies that
$\mbox{e}^{-\eta}\Gamma$ is bounded. A bound on $\mbox{e}^{-\eta}H_t$
follows if we instead let 
$P=\mbox{e}^{-2\eta}\alpha^{-1/2}t^2H_t.$
Using the auxiliary equations (\ref{aux1}-\ref{aux4}) we get
expressions 
on $P_{\theta}$ and $P_t.$ In the expression for $P_t$ we get a term
containing $A_t$ which we treat as in the case of bounding $U,$
i,e. we use a H\"{o}lder argument to bound that term in terms of 
$E(t).$ We also use that $\mbox{e}^{-\eta}\Gamma, A$ 
and $\tilde{\eta}$ are bounded but all ideas have already been used above
so we leave them out and conclude that $\mbox{e}^{-\eta}|H_t|$ is
bounded on $[t_0,t_+)$ as well. 

\textit{Step 2. }(Bounds on $U_{t},U_{\theta},A_{t},
A_{\theta},\eta_{t},\tilde{\eta}_{\theta},
\alpha_{t}$ and $Q(t)$.)\\ 
To bound the derivatives of $U$ we use light-cone estimates in a similar way 
as for the contracting direction. However, the matter terms must be treated 
differently and we need to carry out a careful analysis of the characteristic 
system associated with the Vlasov equation. Let us define 
\begin{eqnarray} 
X&=&\frac{1}{2}(U_{t}^2+\alpha U_{\theta}^2)+\frac{\mbox{e}^{4U}}{8t^2}
(A_{t}^2+\alpha A_{\theta}^2),\label{GE}\\
Y&=&\sqrt{\alpha}U_{t}U_{\theta}+\frac{\mbox{e}^{4U}}{4t^2}A_{t}A_{\theta},
\label{HE} 
\end{eqnarray} 
and 
\begin{eqnarray} 
\chi&=&\frac{1}{\sqrt{2}}(\partial_{t}+\sqrt{\alpha}
\partial_{\theta})\\ 
\zeta&=&\frac{1}{\sqrt{2}}(\partial_{t}-\sqrt{\alpha}
\partial_{\theta}) 
\end{eqnarray} 
A motivation for the introduction of these quantities is based on similar 
arguments as those given in Step 1, Section 4. For details we refer to 
\cite{b1}. 

\textbf{Remark. }We use the same notations, $X$ and $Y$, 
as in the contracting direction, and below we continue to carry over 
the notations. The analysis in the respective direction is independent 
so there should be no risk of confusion.\\ 

By using the evolution equation (\ref{evol2}), a short computation shows that 
\begin{eqnarray} 
\displaystyle\zeta(X+Y)&=&\frac{\alpha_{t}}{2\sqrt{2}\alpha}
(X+Y)
\nonumber\\ 
\displaystyle & &-\frac{1}{\sqrt{2}t}\left(U_{t}^2+\sqrt{\alpha}U_{t}U_{\theta}
+\frac{\mbox{e}^{4U}}{4t^2}(\alpha A_{\theta}^2
+\sqrt{\alpha}A_{t}A_{\theta})\right)
\nonumber\\ 
\displaystyle & &+\frac{(U_{t}+\sqrt{\alpha}U_{\theta})}{2\sqrt{2}}
(\mbox{e}^{4U-2\eta}\Gamma^2+\alpha
\mbox{e}^{2(\eta-U)}\kappa)\nonumber\\
\displaystyle &
&+\frac{(A_{t}+\sqrt{\alpha}A_{\theta})}{4\sqrt{2}}(\mbox{e}^{4U-2\eta}\Gamma
H_t+2\alpha\mbox{e}^{2\eta}S_{23}), 
\nonumber\\ 
\label{han11}\\ 
\displaystyle\chi(X-Y)&=&\frac{\alpha_{t}}{2\sqrt{2}\alpha}
(X-Y)
\nonumber\\ 
\displaystyle & &-\frac{1}{\sqrt{2}t}\left(U_{t}^2-\sqrt{\alpha}U_{t}U_{\theta}
+\frac{\mbox{e}^{4U}}{4t^2}(\alpha A_{\theta}^2
-\sqrt{\alpha}A_{t}A_{\theta})\right)
\nonumber\\ 
\displaystyle & &+\frac{(U_{t}-\sqrt{\alpha}U_{\theta})}{2\sqrt{2}}
(\mbox{e}^{4U-2\eta}\Gamma^2+\alpha
\mbox{e}^{2(\eta-U)}\kappa)\nonumber\\
\displaystyle &
&+\frac{(A_{t}-\sqrt{\alpha}A_{\theta})}{4\sqrt{2}}(\mbox{e}^{4U-2\eta}\Gamma
H_t+2\alpha\mbox{e}^{2\eta}S_{23}),\nonumber\\ 
\label{hanvi} 
\end{eqnarray} 
Here $\kappa=\rho-P_1+P_2-P_3.$ 
Now we wish to integrate these equations along the integral curves of 
the vector fields $\chi$ and $\zeta$ respectively 
(let us henceforth call these integral curves null curves, since they 
are null with respect to the two-dimensional ``base spacetime''). Below 
we show that the quantity 
\begin{equation} 
W(t):=\sup_{\theta\in S^1}X(t,\cdot)+Q^2(t), 
\end{equation} 
is uniformly bounded on $[t_0,t_+)$ by deriving the inequality 
\begin{equation} 
W(t)\leq C+\int_{t_0}^{t}W(s)\ln{W(s)}ds.\label{GlogG} 
\end{equation} 
First we note, exactly as in the contracting direction, that the
symmetry 
implies 
that 
$$V^{2}(t)\mbox{e}^{U(t,\Theta(t))}$$ and 
$$V^{2}(t)A\mbox{e}^{U}+V^{3}(t)t\mbox{e}^{-U(t,\Theta(t))},$$ 
are conserved. Here $V^{2}(t), V^{3}(t)$ and $\Theta(t)$ are solutions 
to the characteristic system associated to the Vlasov equation. From Step 2 
we have that $U$ and $A$ are uniformly bounded on $[t_0,t_+)$. 
Hence $|V^2(t)|$ 
and $|V^3(t)|$ are both uniformly bounded on $[t_0,t_+)$, and since 
the initial distribution function $f_0$ has compact support we conclude that 
\begin{equation} 
\sup\{|v^2|+|v^3|:\exists (s,\theta)
\in [t_0,t]\times S^1\mbox{ with }f(s,\theta,v)\not= 0\}, 
\end{equation} 
is uniformly bounded on $[t_0,t_+)$. Therefore, in order to control 
$Q(t)$ it is sufficient to control 
\begin{equation} 
Q^1(t):=\mbox{sup}\{|v^1|:\exists (s,\theta)\in [t_0,t]\times 
S^1\mbox{such that} f(s,\theta,v)\not= 0\}.\label{Q1} 
\end{equation} 
Below we introduce the uniformly bounded function $\gamma(t)$ to denote 
estimates regarding the variables $v^2$ and $v^3$. As observed in \cite{a} 
there is some cancellation to take advantage of in the matter term 
$(\rho-P_1)$ which appears in the equations for $X+Y$ and $X-Y$ above. 
It is proved there that 
\begin{equation}
(\rho-P_1)(t,\theta)\le C\gamma(t)\ln{Q^1(t)}
\end{equation}
In a similar fashion we can estimate $P_2,P_3$ and $S_{23}$. 

Let us now derive (\ref{GlogG}). As in Step 2 in Section 4 we integrate 
the equations above for $X+Y$ and $X-Y$ along null paths. For 
$t\geq t_{0}$ integrate along the two null paths defined by 
$\chi$ and 
$\zeta$, starting at $(t_0,\theta)$ and add the results. In
this way the following inequality can be derived:
\begin{equation}\label{supg}
\displaystyle\sup_{\theta}X(t,\cdot)\le C+C(t)\int_{t_0}^{t}
[1+\sup_{\theta}X(s,\cdot)]\ln{Q^1(s)}]ds
\end{equation}
In doing this it is important to use the fact that 
$\alpha e^{2\eta}=e^{2\tilde\eta}$ and the above estimates for matter
quantities in terms of $\ln Q^1$. What is new compared to \cite{a} is the
occurrence of twist quantities and these can be treated using the fact
that $e^{-\eta}\Gamma$ and $e^{-\eta}H_t$ are bounded.  

Let us now derive an estimate for $Q^1$ in terms of $\sup_{\theta}X$. 
\begin{lemma} 
Let $Q^1(t)$ and $X(t,\theta)$ be as above. Then 
\begin{equation} 
\displaystyle |Q^1(t)|^2\leq C+
D(t)\int_{t_0}^{t}[(Q^1(s))^2+\sup_{\theta}X(s,\cdot)]ds, 
\end{equation} 
where $C$ is a constant and $D(t)$ is a uniformly bounded function 
on $[t_0,t_+)$. 
\end{lemma} 
\textit{Proof. }The characteristic equation for $V^1$ associated to the 
Vlasov equation and the constraint equations (\ref{constr1}) and 
(\ref{constr2}) imply that 
\begin{equation} 
\frac{d}{ds}(V^1(s))^2=2V^1(s)\frac{d}{ds}V^1(s)=T_1+T_2+T_3+T_4, 
\end{equation} 
Here $T_1$, $T_2$ and $T_3$ are expressions given in \cite{a} and they can
be estimated just as in the corresponding lemma of that reference. The
additional term $T_4$, which collects together the contributions involving
the twist quantities, is
\begin{equation}
T_4=-2s(V^1)^2\frac{e^{-2\eta}}{4}(e^{4U}\Gamma^2+s^2H_t^2)
-2V^1(s)[\mbox{e}^{-\eta}(\mbox{e}^{2U}\Gamma V^2+sH_{t}V^3)]
\end{equation}
It is easily seen that $|T_4|\le C\gamma(t)(Q^1(t))^2$ and this suffices 
to obtain the desired estimate.
\begin{flushright} 
$\Box$ 
\end{flushright} 
Combining the estimate for $(Q^1(t))^2$ in the lemma and the estimate 
(\ref{supg}) for $\sup_{\theta}X(t,\cdot)$, we find that $W(t)$ 
satisfies the estimate (\ref{GlogG}) and is thus uniformly bounded. 
The constraint equation (\ref{constr1}) now immediately shows that 
$|\eta_{t}|$ is bounded by
$$2tX+\frac{t\mbox{e}^{-2\eta}}{4}(\mbox{e}^{4U}\Gamma^2+t^2H_{t}^2)
+ t\mbox{e}^{2(\tilde{\eta}-U)}\rho
\leq C(t)[1+\sup_{\theta}X(t,\cdot)+(Q(t))^3],$$ 
since $$\rho=\int_{\mathbb{R}^3}fdv\leq
\|f_0\|_{\infty}\int_{|v|\leq Q(t)}dv\leq C(Q(t))^3.$$ Recall that
$\mbox{e}^{-2\eta}(\mbox{e}^{4U}\Gamma^2+t^2H_{t}^2)$ is known to
be bounded. Thus also $\eta$ is bounded which implies that both
$|\Gamma|$ and $|H_t|$ are bounded. The bound on 
$\eta$ also provides a bound on
$|\tilde{\eta}_{\theta}|$ using the constraint equation (\ref{constr2}),
where $|J_1|$ is estimated in terms of $Q(t)$. 
Analogous arguments show that $|\alpha_{t}|$ is uniformly bounded. 
The uniform bound on $X$ provides bounds on $|U_t|$ and $|A_{t}|$, but 
to conclude that $|U_{\theta}|$ and $|A_{\theta}|$ are bounded we have to 
show that $\alpha$ 
stays uniformly bounded away from zero. Equation (\ref{constr3}) is easily 
solved, 
\begin{equation} 
\displaystyle \alpha(t,\theta)=\alpha(t_0,\theta)
\mbox{e}^{\int_{t_0}^tF(s,\theta)ds}, 
\end{equation} 
where $$F(t,\theta):=-2t\mbox{e}^{2(\tilde{\eta}-U)}(\rho-P_1)
-t\mbox{e}^{-2\eta}(\mbox{e}^{4U}\Gamma^2+t^2H_{t}^2),$$ 
which is uniformly bounded from below. 
Hence $|U_{\theta}|$ and $|A_{\theta}|$ are bounded 
and Step 2 is complete. 

\textit{Step 3. }(Bounds on $\partial f$, $\alpha_{\theta}$ 
and $\eta_{\theta}$.)\\ 
The main goal in this step is to show that the first derivatives 
of the distribution function are 
bounded. In view of the bound on $Q(t)$ we then also obtain bounds 
on the first derivatives of the matter terms $\rho, J_k, S_{jk}$ and 
$P_k,$ $j,k=1,2,3;j\ne k$. 
Such bounds together with bounds on the $\theta$ derivatives of the twist
quantities almost immediately lead to bounds on $\alpha_{\theta}$ 
and $\eta_{\theta}$. 

Recall that the solution $f$ can be written in the form 
\begin{equation} 
f(t,\theta,v)=f_{0}(\Theta(0,t,\theta,v),V(0,t,\theta,v)),\label{solu1} 
\end{equation} 
where $\Theta(s,t,\theta,v), V(s,t,\theta,v)$ is the solution 
to the characteristic system
\begin{eqnarray} 
\displaystyle\frac{d\Theta}{ds}&=&\sqrt{\alpha}\frac{V^1}{V^0}, 
\label{chtta}\\ 
\displaystyle\frac{dV^1(s)}{ds}&=&-(\eta_{\theta}-U_{\theta}+
\frac{\alpha_{\theta}}{2\alpha})\sqrt{\alpha}V^{0}-(\eta_{t}-U_{t})V^{1} 
\nonumber\\ 
\displaystyle & &-\frac{\sqrt{\alpha}U_{\theta}}{V^{0}}
((V^{2})^{2}-(V^{3})^{2})
+\frac{\sqrt{\alpha}A_{\theta}}{sV^0}\mbox{e}^{2U}V^2V^3\nonumber\\
\displaystyle & &-\mbox{e}^{-\eta}(\mbox{e}^{2U}\Gamma V^2+sH_{t}V^3),
\label{chdva1}\\
\displaystyle\frac{dV^2}{ds}&=&-U_{t}V^{2}-\sqrt{\alpha}U_{\theta}
\frac{V^{1}V^{2}}{V^{0}},
\label{chdv2}\\ 
\displaystyle\frac{dV^3}{ds}&=&-(\frac{1}{s}-U_{t})V^{3}
+\sqrt{\alpha}U_{\theta}\frac{V^{1}V^{3}}{V^{0}}
\nonumber\\ 
\displaystyle & &-\frac{\mbox{e}^{2U}}{s}(A_{t}
+\sqrt{\alpha}A_{\theta}\frac{V^1}{V^0})V^2,
\label{chdv3} 
\end{eqnarray} 
with the property 
$\Theta(t,t,\theta,v)=\theta$, $V(t,t,\theta,v)=v$. 
Hence, in order to establish bounds on the first derivatives of $f$ it is 
sufficient to bound $\partial\Theta$ and $\partial V$ since $f_0$ is smooth. 
Here $\partial$ denotes the first order derivative with respect 
to $t,\theta$ or $v$. 
Evolution equations for $\partial\Theta$ and $\partial V$ are provided 
by the characteristic system above. However, the right hand sides will 
contain second order derivatives of the field components, but so far 
we have only obtained bounds on the first order derivatives 
(except for $\eta_{\theta}, \alpha_{\theta}$). 
Yet, certain combinations of second order derivatives can 
be controlled. Behind this observation lies 
a geometrical idea which plays a fundamental role in 
general relativity. An important property of curvature is its control 
over the relative behaviour of nearby geodesics. Let $\gamma(u,\lambda)$ be 
a two-parameter family of geodesics, i.e. for each fixed $\lambda$, 
the curve $u\mapsto\gamma(u,\lambda)$ is a geodesic. 
Define the variation vector field 
$Y:=\gamma_{\lambda}(u,0)$. This vector field satisfies 
the geodesic deviation equation (or Jacobi equation) (see eg. \cite{h1}) 
\begin{equation} 
\frac{D^2Y}{Du^2}=R_{Y\gamma'}\gamma',\label{ged} 
\end{equation} 
where $D/Du$ is the covariant derivative, $R$ the Riemann curvature tensor, 
and $\gamma':=\gamma_{u}(u,0)$. 
Now, the Einstein tensor is closely related to the curvature tensor and 
since the Einstein tensor is proportional to the energy momentum tensor 
which we can control from Step 2, it is meaningful, in view of (\ref{ged}) 
(with $Y=\partial \Theta$), 
to look for linear combinations of $\partial\Theta$ and $\partial V$ which 
satisfy an equation with bounded coefficients. More precisely, we want to 
substitute the twice differentiated field components which appear by 
taking the derivative of the characteristic system by using 
the Einstein equations. 
The geodesic deviation equation has previously played an important role 
in studies of the Einstein-Vlasov system (\cite{r2}, \cite{r1} and 
\cite{r5}). 
\begin{lemma} 
Let $\Theta(s)=\Theta(s,t,\theta,v)$ and $V^k(s)=V^k(s,t,\theta,v)$, 
$k=1,2,3$ be a solution to the characteristic system 
(\ref{chtta})-(\ref{chdv3}). Let $\partial$ denote 
$\partial_{t},\partial_{\theta}$ 
or $\partial_{v}$, and define 
\begin{eqnarray} 
\displaystyle\Psi&=&\alpha^{-1/2}\partial\Theta,\label{Psi}\\ 
\displaystyle Z^1&=&\partial V^1+
\left(\frac{\eta_tV^0}{\sqrt{\alpha}}-\frac{U_tV^0}
{\sqrt{\alpha}}\,\frac{(V^0)^2-(V^1)^2+(V^2)^2-(V^3)^2}{(V^0)^2-(V^1)^2}
\right.\nonumber\\ 
\displaystyle & &\left.+\,U_{\theta}\,
\frac{V^1((V^2)^2-(V^3)^2)}{(V^0)^2-(V^1)^2}-\frac{A_{t}\mbox{e}^{2U}}
{\sqrt{\alpha}t}\frac{V^0V^2V^3}{(V^0)^2-(V^1)^2}\right. 
\nonumber\\ 
\displaystyle & & \left.+A_{\theta}
\frac{V^1V^2V^3}{(V^0)^2-(V^1)^2}\right)\partial\Theta,
\label{Z1}\\ 
\displaystyle Z^2&=&\partial V^2+V^2U_{\theta}\,\partial\Theta,
\label{Z2}\\ 
\displaystyle Z^3&=& 
\partial V^3-(V^3U_{\theta}-\frac{\mbox{e}^{2U}}{s}V^2A_{\theta})\,
\partial\Theta.\label{Z3} 
\end{eqnarray} 
Then there is a matrix $A=\{a_{lm}\}$, $l,m=0,1,2,3$, 
such that $$\Omega:=(\Psi,Z^1,Z^2,Z^3)^T$$ satisfies 
\begin{equation} 
\displaystyle\frac{d\Omega}{ds}=A\Omega,\label{Omega} 
\end{equation} 
and the matrix elements $a_{lm}=a_{lm}(s,\Theta(s),V^k(s))$ are 
all uniformly bounded on $[t_0,t_+)$. 
\end{lemma} 
\textit{Sketch of proof. }Once the ansatz 
(\ref{Psi})-(\ref{Z3}) has been found this is only a lengthy calculation. 
To illustrate the type of calculations 
involved we show the easiest case, i.e. the $Z^2$ term. 
\begin{eqnarray} 
\displaystyle\frac{dZ^2}{ds}&=&\frac{d}{ds}(\partial V^2+V^2U_{\theta}
\partial\Theta)\nonumber\\
\displaystyle&=&\partial(\frac{d}{ds}V^2)
+\frac{dV^2}{ds}U_{\theta}\partial\Theta\nonumber\\ 
\displaystyle&\phantom{h}&+V^2(U_{t\theta}+U_{\theta\theta}\frac{d\Theta}{ds})
\partial\Theta+V^2U_{\theta}\partial(\frac{d\Theta}{ds}).\nonumber\\ 
\end{eqnarray} 
Now we use (\ref{chtta}) and (\ref{chdv2}) to substitute for $d\Theta/ds$ 
and $dV^2/ds$. We find that the right hand side equals 
\begin{eqnarray} 
&\displaystyle\partial(-U_tV^2-\sqrt{\alpha}U_{\theta}\frac{V^1V^2}{V^0})+
(-U_tV^2-\sqrt{\alpha}U_{\theta}\frac{V^1V^2}{V^0})U_{\theta}\partial\Theta&
\nonumber\\ 
&\displaystyle +V^2(U_{t\theta}+U_{\theta\theta}\sqrt{\alpha}\frac{V^1}{V^0})
\partial\Theta+V^2U_{\theta}\left(\frac{\alpha_{\theta}V^1}{2\sqrt{\alpha}V^0}
\partial\Theta+\sqrt{\alpha}\partial(\frac{V^1}{V^0})\right)&. 
\nonumber
\end{eqnarray} 
Taking the $\partial$ derivative of the first term we find that all 
terms of second order derivatives and terms containing $\alpha_{\theta}$ 
cancel. Next, since 
\begin{equation} 
\displaystyle -\sqrt{\alpha}U_{\theta}\partial\left(\frac{V^1V^2}{V^0}\right)
+\sqrt{\alpha}U_{\theta}V^2\partial\left(\frac{V^1}{V^0}\right)=-\sqrt{\alpha}
U_{\theta}\frac{V^1}{V^0}\partial V^2, 
\end{equation} 
we are left with 
\begin{equation} 
\displaystyle \frac{dZ^2}{ds}=-(U_tV^2+\sqrt{\alpha}
U_{\theta}\frac{V^1V^2}{V^0})U_{\theta}\partial\Theta-(U_t+\sqrt{\alpha}
U_{\theta}\frac{V^1}{V^0})\partial V^2.\label{z2b} 
\end{equation} 
Finally we express this in terms of $\Psi, Z^1,Z^2$ and $Z^3$. Here 
this is easy and we immediately get 
\begin{displaymath} 
\frac{dZ^2}{ds}=-(U_{t}+\sqrt{\alpha}U_{\theta}\frac{V^1}{V^0})Z^2. 
\end{displaymath} 
Clearly, the map $(\partial\Theta,\partial V^k)\mapsto (\Psi,Z^k)$ is 
invertible so that this step is easy also in the other cases. 
It follows that the matrix elements $a_{2m}$, $m=0,1,2,3$, are 
uniformly bounded on $[t_0,t_+)$ (only $a_{22}$ is nonzero here). 
The computations for the other terms are similar. For the $Z^1$ term we point 
out that the evolution equations (\ref{evol2}) and (\ref{evol4}) should be 
invoked and that the matrix element $a_{10}$ contains $\eta_{\theta}$ 
and $\alpha_{\theta}/2\alpha$, but they combine and form
$\tilde{\eta}_\theta,$ and that derivatives of $\mbox{e}^{-\eta}\Gamma$ and
$\mbox{e}^{-\eta}H_t$ appear. The latter terms are easily seen to be
bounded in view of the auxiliary equations. For example, from (\ref{aux1})
we get a bound on $\partial_{\theta}(\mbox{e}^{-\eta}\Gamma),$ 
$$\frac{\partial}{\partial\theta}(\mbox{e}^{-\eta}\Gamma)=-2
\mbox{e}^{\eta+\tilde{\eta}-4U}J_{2}
-\mbox{e}^{\tilde{\eta}-\eta-4U}\Gamma\frac{\partial}
{\partial\theta}(\mbox{e}^{-\tilde{\eta}+4U}).$$ 
The right hand side is bounded since $U, U_{\theta}, \eta, \tilde{\eta}, 
\partial_{\theta}\tilde{\eta},$ and $\Gamma$ are all bounded as was
shown in Step 1 and 2. 
\begin{flushright} 
$\Box$ 
\end{flushright} 
From the lemma it now immediately follows that $|\Omega|$ is 
uniformly bounded on $[t_0,t_+)$. Moreover, 
since the system (\ref{Psi})-(\ref{Z3}) is invertible with 
uniformly bounded coefficients 
we also have uniform bounds 
on $|\partial\Theta|$ and $|\partial V^k|$, $k=1,2,3$. 
In view 
of the discussion at the beginning of this section we see that 
the distribution function $f$ and the matter quantities 
$\rho, J_k, S_{jk}$ and $P_k$, are all uniformly $C^1$ bounded. 
From the constraint equation (\ref{constr3}) we now obtain a uniform 
bound on $\alpha_{\theta}$ by a simple Gronwall argument using as usual 
the identity $\alpha\mbox{e}^{2(\eta-U)}=\mbox{e}^{2(\tilde{\eta}-U)}$. 
Finally this yields a uniform bound on $\eta_{\theta}$ since 
$$\eta_{\theta}=\tilde{\eta}_{\theta}-\frac{\alpha_{\theta}}{2\alpha}$$ 
and $\alpha$ stays uniformly bounded away from zero. 

\textit{Step 4. }(Bounds on second and higher 
order derivatives.)\\ 
It is now easy to obtain bounds on second order derivatives on $U$ and $A$ 
by using light cone arguments. We define $X$ and $Y$ by 

\begin{eqnarray} 
X&=&\frac{1}{2}(U_{tt}^2+\alpha U_{t\theta}^2)+\frac{\mbox{e}^{4U}}{8t^2}
(A_{tt}^2+\alpha A_{t\theta}^2),\label{GEt}\\
Y&=&\sqrt{\alpha}U_{tt}U_{t\theta}+\frac{\mbox{e}^{4U}}{4t^2}A_{tt}A_{t\theta},
\label{HEt} 
\end{eqnarray} 
and use the differentiated (with respect to $t$) evolution equations for $U$ 
and $A$ to obtain equations similar to (\ref{han11}) and (\ref{hanvi}). 
In this case a straightforward light cone argument applies since we have 
control of the differentiated matter terms. $U_{\theta\theta}$ 
and $A_{\theta\theta}$ are then 
uniformly bounded in view of the evolution equations 
(\ref{evol2}) and (\ref{evol4}). 
Bounds on second order derivatives on $f$ then follows from (\ref{Omega}) by 
studying the equation for $\partial\Omega$. The only thing to notice is 
that $\tilde{\eta}_{\theta\theta}$ is controlled by (\ref{constr2}). 
It is clear that this reasoning can be continued to give uniform bounds 
on $[t_0,t_+)$ for higher order derivatives as well. 
\begin{flushright} 
$\Box$ 
\end{flushright} 

\section{Proofs of the main theorems}\label{proof}

In this section the analytical and geometrical information obtained in
previous sections is combined to obtain the main results of the 
paper. 

\noindent{\em Proof of Theorem~\ref{theorem2}.}
For a spacetime satisfying the hypotheses of the theorem we
know from Section~\ref{t2} that a conformal coordinate system
can be introduced on a neighbourhood of the initial hypersurface $S_0$
corresponding to the original data. The direct analogues of the results of
Section 5 of \cite{arr} hold and can be proved by the same arguments.
In fact the situation is slightly simpler since in the case of $T^2$
symmetry there is an obvious choice of two Killing vectors while in
\cite{arr} it was necessary to worry about choosing two from a total
of three Killing vectors in an appropriate way. Cf. also \cite{b1}
where this type of argument was introduced for the first time.
By these results it follows that the region where the solution exists
can be extended to the past so as to include a Cauchy surface $S_A$ of 
constant areal time. Moreover, either the conformal time coordinate extends 
to all negative values, or $R$ tends to 
zero as the past boundary of the region covered by conformal
coordinates is approached. In the first of these cases the region
covered by the conformal time coordinate includes the entire past of the
initial hypersurface in the maximal Cauchy development, as follows from
the arguments of Section 5 of \cite{arr}. Also the past of $S_A$ in that  
region admits a foliation by hypersurfaces of constant $R$. In that region 
we can transform to areal coordinates. For we can choose a new spatial 
coordinate $\theta$ so that its coordinate lines in $Q$ are orthogonal to 
that foliation. In the second case (where $R$ tends to zero on the 
boundary of the region covered by conformal coordinates) the past of 
$S_A$ is also covered by areal coordinates. It exhausts the past of $S_A$ 
in the maximal Cauchy development, as will now be shown. If the spacetime
could be extended to the past there would be a sequence of points tending
to the boundary of the original spacetime in the extension. Along this
sequence $R$ would have to tend to zero. However the function $R$ which
is globally defined on the maximal Cauchy development must tend to a
non-zero limit along the sequence approaching a point of the
extension. Thus the existence of an extension leads to a contradiction.
It follows that in both cases the entire past of $S_0$ 
in the maximal Cauchy development is covered. As a consequence of the results 
of Section~\ref{expand} the spacetime and the areal time coordinate 
can be extended so that the time coordinate covers the interval 
$(R_0,\infty)$. It can then be concluded by the argument at the end of 
Section 5 of \cite{arr} that the entire future of $S_0$ in the maximal Cauchy 
development is covered.
\begin{flushright} 
$\Box$ 
\end{flushright}

\noindent{\em Proof of Theorem~\ref{theorem1}.}
The mean curvature of the hypersurfaces of constant areal time is 
\begin{equation}\label{negative}
\tr k=-e^{-\eta+U}\alpha^{-1/2}(\eta_t-U_t+t^{-1})
\end{equation} 
From the field equations it follows that
\begin{equation}
\eta_t-U_t+t^{-1}\ge tU_t^2-U_t+t^{-1}= \frac{3}{4}tU_t^2+t(\frac{1}{2}
U_t-t^{-1})^2
\end{equation}
Hence there are Cauchy surfaces with everywhere negative mean 
curvature. Under these circumstances it
was shown by Henkel \cite{henkel} that the initial singularity is
a crushing singularity and thus a neighbourhood of it can be foliated by 
CMC hypersurfaces. Given one CMC hypersurface the statement in Theorem 1 
about the range 
of the CMC time coordinate follows from \cite{r3}. It remains to see that 
the CMC foliation covers the entire future of the initial hypersurface. This 
can be proved by an argument used in the case of hyperbolic symmetry in
\cite{arr} which will now be recalled. It is enough to show that if 
$p$ is any point of the spacetime there is a compact CMC hypersurface 
which contains $p$ in its past. Let $S_1$ be the Cauchy surface of constant 
areal time passing through $p$. Equation (\ref{negative}) shows that the 
mean curvature of $S_1$ is strictly negative. Hence it has a maximum value
$H_1<0$. Let $S_2$ be the compact CMC hypersurface with mean curvature
$H_1/2$. Then the infimum of the mean curvature of $S_2$ is greater
than the supremum of the mean curvature of $S_1$ and a standard argument
\cite{marsden} shows that $S_2$ is strictly to the future of $S_1$. Hence 
$p$ is in the past of $S_2$, as required.
\begin{flushright} 
$\Box$ 
\end{flushright}


\end{document}